# Air nonlinear dynamics initiated by ultra-intense lambda-cubic THz pulses

Mostafa Shalaby[1] and Christoph P. Hauri[1,2]

[1]Paul Scherrer Institute, SwissFEL, 5232 Villigen PSI, Switzerland.

[2]École Polytechnique Fédérale de Lausanne, 1015 Lausanne, Switzerland.

*Correspondence to: most.shalaby@gmail.com and christoph.hauri@psi.ch

**Abstract**: We report on the measurement of the instantaneous Kerr nonlinearity and the retarded alignment of air molecules $CO_2$, $N_2$ and $O_2$ triggered by an intense, lambda-cubic terahertz pulse. The strong-field, impulsive low-frequency excitation (4 THz) leads to field-free alignment dynamics of these molecules thanks to the THz-induced transient dipole moments in the otherwise non-polar molecules. The strong coupling to the terahertz electric transient results in the excitation of coherent long-living rotational states at ambient pressure. Beyond fundamental investigations of nonlinear properties in gases, our results suggest a novel route towards field-free molecular alignment at laser intensity well below the ionization threshold.

Air turns into a nonlinear medium for electromagnetic waves under exceptionally strong fields that manifest itself in phenomena like the instantaneous Kerr nonlinearities and retarded molecular alignment [1]. Up to present, the minuscule nonlinear response of air has limited the exploration of its nonlinear properties to the optical frequency regime owing to the availability of intense near-infrared (nIR) lasers (e.g Ti:sapphire with a carrier frequency ≈380 THz). From a fundamental science point of view nonlinear excitation of air (and other gases) by pulses with a lower carrier frequency, in the so-called *terahertz gap* (1-10 THz), is of large interest as the dominating quantities, like the Kerr nonlinear coefficient ($n_2$) are yet unknown. The determination of the air Kerr nonlinear coefficient is essential for prediction and understanding novel nonlinear phenomena [2] such as terahertz (THz) self-guiding (filamentation) and THz self-phase modulation. Furthermore, such field transients at frequencies in the THz gap match the natural timescale of rotational molecular dynamics which hold promise for enhanced control on molecular alignment.

Up to now both the observation of Kerr nonlinearities and molecular alignment in air at THz frequencies was hampered by the technological void of intense sources [3-9]. Field-free alignment/orientation of molecules is conventionally done by nIR laser technology which allows for preparing the molecules in a preferred angular distribution prior to their interrogation [10,11 and references therein]. The elimination of a randomly oriented molecule distribution is beneficial for numerous applications including high harmonic generation [12-14], THz emission [15, 16], the control of chemical, surface [17] and photoreactions whose rates depend on the molecular orientation, and others [18-20]. The two mostly applied alignment techniques are based on intense nIR femtosecond laser pulses where alignment is achieved by means of molecular two-color excitation [21] or through the formation of elongated plasma channels (filaments) [22]. Both of these alignment procedures require typically high laser pulse intensities ($\geq 10^{13}$ W/cm$^2$) which may cause unwanted ionization and excitation of electronic-vibronic states during the interaction. Molecular orientation/alignment by a THz pulse offers in principle superior conditions as the THz photon quantum energy is 2-3 orders of magnitude smaller than the nIR quantum (meV vs eV). This considerably reduces the probability of multi-photon ionization. However, molecular alignment of non-polar molecules, such as air, has never been demonstrated by means of THz radiation.

In this work, we report two advances. First, the quantification of the nonlinear Kerr coefficient $n_2$, measured at THz frequencies and derived from the instantaneous Kerr nonlinearity in air. Second, the demonstration of coherent molecular alignment dynamics of the *non-polar* nitrogen $N_2$, oxygen $O_2$, and carbon dioxide $CO_2$ in dry air at ambient temperature and pressure by a strong THz transient. As these molecules do not carry a permanent dipole, the key for the observed alignment dynamics at low THz frequencies is the recently developed ultra-intense lambda-cubic THz bullets [23]. The three orders of magnitude higher THz field than in previous THz-gas experiments [24-26] enables the induction of a transient dipole moment in the otherwise non-polar molecules leading to a strong transient coupling with the THz electric field component.

In order to measure the THz-induced nonlinearity, we performed THz pump-optical probe spectroscopy (Fig. 1(a)). Our THz trigger consists of the $\lambda^3$ THz bullet (i.e. a diffraction-limited and transform-limited single-cycle pulse) with maximum electric field strengths of 33 MV/cm at 3.9 THz center frequency [23]. The corresponding peak intensity is 1.44 TW/cm$^2$. The probe pulses are 800 nm-centered pulses polarized at $45^0$ with respect to the $x$-polarized THz pump.

The considered air molecules have zero permanent dipole moment and a nearly vanishing average linear susceptibility $\chi^{(1)}$ of 5.4 x 10$^{-4}$. In gases, the third order susceptibility $\chi^{(3)}$ is the first nonlinear susceptibility term which is present, due to symmetry constraints. $\chi^{(3)}$ of air is extremely weak, being 5.4 x $10^{-25}$ m$^2$/V$^2$ [27]. Observation of nonlinearities in air requires therefore intensities on the order of $10^{11}$ W/cm$^2$. Up to now such intensities were not available in the THz gap but rather at optical frequencies. At optical frequencies, a multitude of nonlinear phenomena [28-30] are observed as consequence of coupling the laser electric field to the polarization through $\chi^{(3)}$ and similar phenomena are, in principle, expected at THz frequencies.

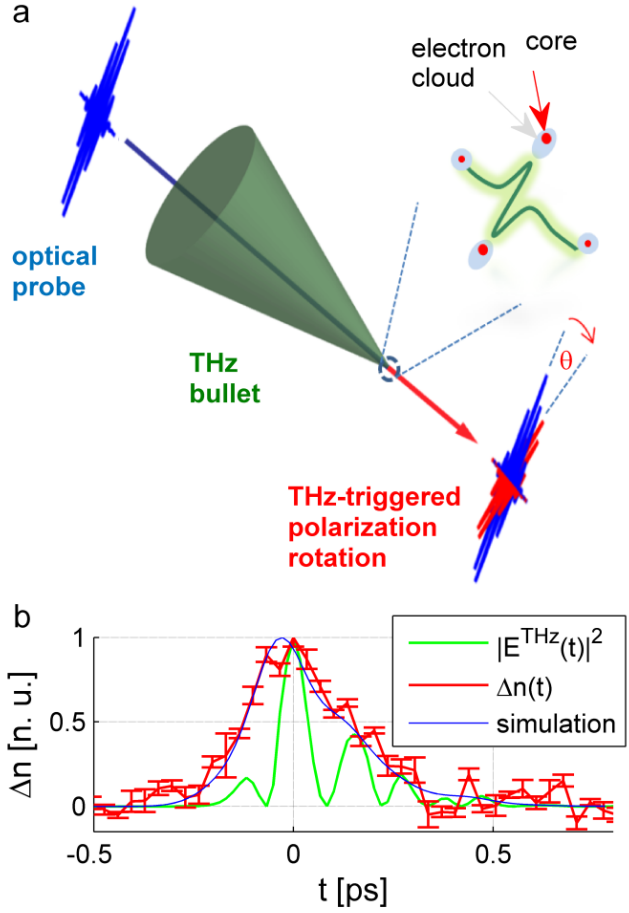

FIG. 1. (color online). (a) Schematic diagram of the measurement configuration of the THz-induced nonlinearity. An intense THz bullet nonlinearly induces a transient dipole moment by deforming (displacing) the electron clouds in the air molecules along the ($x$-) polarization direction of the laser. The near infrared (NIR) probe is initially polarized at $45^0$ from the THz polarization. The probe experiences THz-induced birefringence in air. (b) A comparison between the THz intensity, the measured birefringence $\Delta n = n_x - n_y$ and simulations.

A change in the electronic susceptibility can be experimentally determined by measuring the refractive index $n$. In a centro-symmetric material, they are related [28, 29] by

$$n = (\chi + 1)^{1/2} = \left(\chi^{(1)} + 1\right)^{1/2} + \frac{3\chi^{(3)}}{4n_0}|E|^2/z_0 = n_0 + n_2 I \qquad (1)$$

where $n_0$ and $I$ are the linear refractive index and intensity of the exciting field. $n_2$ is the nonlinear refractive index, commonly referred to as Kerr coefficient.

The induced birefringence $\Delta n = n - n_0 = n_2 I$ is determined from the measurement of the THz induced phase retardation of the probe $\Delta \varphi = \Delta n \, \omega \, L/c$ where $\omega$ and $c$ are the optical frequency of the probe and the speed of light in vacuum, respectively [30]. $I$ is the THz intensity. $L$ is the interaction length. For a tightly focused laser pulse, like our THz bullet, the THz intensity changes significantly around the focus. It is thus more accurate to estimate the induced birefringence from $\Delta\varphi = \omega/c \int I(z)dz$ where $z$ is the propagation direction. As our probe (optical) pulse duration is comparable to the pump (THz) pulse duration, the detected birefringence is a convolution between the THz intensity profile and optical intensity profile. Figure 1(b) shows the estimated $\Delta n$ in comparison with $I$ where $\Delta n$ is shown to follow the convolution between $I$ and a Gaussian probe pulse intensity profile, confirming the Kerr-like instantaneous nonlinearity in air. The maximum change of refractive index is 1.12 x$10^{-7}$ corresponding to a reduction of the phase velocity of the optical pulse by 34 m/s induced by the THz bullet. As the dispersion of our medium is negligible, the group velocity is reduced by the same amount.

Instantaneous Kerr phase retardation scales linearly with the excitation intensity. However, at extreme intensities, higher order nonlinear Kerr effects start to play a role, reversing the sign of the refractive index change [31, 32]. Such effects manifest themselves in the temporal and spectral responses as well as in violation of the linearity in $\Delta n = n_2 I$. We verified that the excitation levels used here correspond to the linear Kerr regime by measuring the fluence dependence of the induced phase retardation. Figure 2(a) shows the extracted phase retardation for various excitation levels down to 6.9% of the maximum intensity. The time traces show no fluence-dependence change in the temporal response. The maximum phase retardation is shown in Fig. 2(b) versus the excitation field and manifests itself as a linear function of the THz peak intensity.

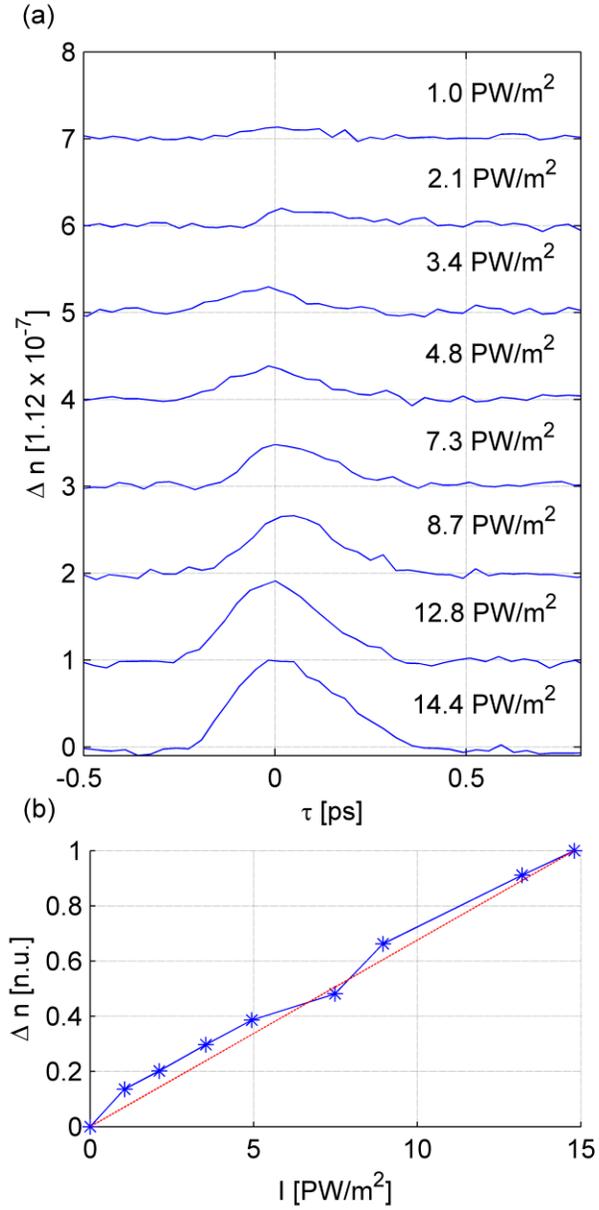

FIG. 2. (color online). (a) The dependence of the measured birefringence on the triggering THz peak intensity. (b) The birefringence is plotted against the THz peak fields (blue). A linear fit response (red) is also shown confirming $\Delta n = n_2 I^{\text{THz}} = n_2 |E^{\text{THz}}|/2z_0$.

We next evaluate the Kerr nonlinear coefficient $n_2$. In comparison with the optical regime, a single cycle THz pulse carries a very broad spectrum of several octaves. As the considered electronic Kerr effect is instantaneous on the THz time scale and the dispersion of our nonlinear medium (dry air) is negligible, the nonlinear response (induced birefringence) is expected to be independent of the excitation frequency.

In order to estimate which part of the spectrum contributes the most to the measured nonlinearity, we employ a set of low pass filters with cut off frequencies ($f_{\text{cutoff}}$) at {18; 9; 6; 3} THz. The peak electric field and intensities was 3.3 GV/m and 1.44 TW/cm$^2$, respectively. The THz intensity profiles from these filters are shown in Fig 3(a). The original (unfiltered) amplitude spectrum of the exciting pulse is shown in Fig. 3(b). Figure 3(c) shows the measured $\Delta n$ for different filters. In the case of $f_{\text{cutoff}} = 3$ THz, the field intensity leads to rotation below the sensitivity of our detection (Fig. 2(a)). Although the peak intensity obtained with the <6 THz LPF was 33% of the total intensity, the measured phase retardation was 74% of the maximum phase retardation which is attributed to the shorter Rayleigh length of high frequency components and the insensitivity of our relatively long probe pulse and spatially large probe size (75 fs and 35 μm (FWHM)) to high frequencies components. For simplicity, in the calculation of the Kerr coefficient, we considered only the spectral contents below 6 THz. From the measured phase retardation and assuming Gaussian beam characteristics around the focus, the Kerr nonlinear coefficient can thus be deduced as 1.3 x 10$^{-23}$ m$^2$/W in the considered THz range.

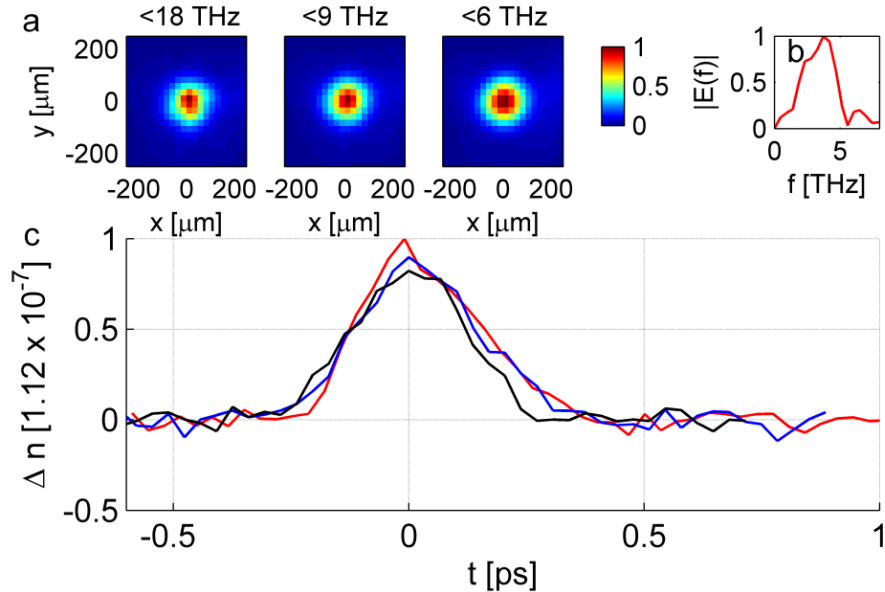

FIG. 3. (color online). (a) THz intensity spot size at the focus for low pass filters with cut off frequency at 18 THz, 9 THz, and 6 THz, respectively. The corresponding average full width at half maxima are 85 μm, 104 μm, and 117 μm. (b) The amplitude spectra of the exciting THz pulse before filtering. The peak spectral density occurs at 3.9 THz. (c) Frequency-dependent contribution of THz induced birefringence in air. Red, blue, and black lines correspond to the measurements done with low-pass filters cutting at 18 THz, 9THz, and 6 THz, respectively.

In addition to the instantaneous effect, we observed evidence of retarded molecular alignment dynamics. Molecular orientation in gases induced by a THz pulse has been observed only in polar molecules (such as $N_2O$ [26] and OCS [24, 25]). Such molecules carry large permanent dipole moments (for example, 0.71 Debye for OCS) which couple to an electromagnetic field via rotational transition frequencies. The orientation can thus be observed at relatively weak THz fields (≈20 kV/cm for OCS). For non-polar molecules, however, much higher field strength (>>MV/cm) is required for alignment which could not be provided by THz sources in the past. Indeed, the established method for aligning molecules with a zero permanent dipole moment is based on intense optical femtosecond pulses (Ti:sapphire) which were confined to an elongated plasma channel (filament) for inducing coherent molecular rotations. In these filaments, however, the alignment pulse undergoes strong nonlinearities and spatial self-confinement upon propagation and the molecules to be aligned are exposed to a laser peak power of typically $5 \times 10^{13}$ $W/cm^2$. Such high intensities can provoke unwanted excitation and damage (particularly for biological samples) prior to the alignment and interrogation [33]. Depending on the gas species under examination the process of filamentation may even require higher intensities, e.g. larger than $10^{14}$ $W/cm^2$. The realization of an advanced technique for aligning non-polar molecules using a non-ionizing laser pulse at much lower intensity (i.e. $\leq 10^{12}$ $W/cm^2$) is thus an important step towards advanced field-free interrogation of molecules. Here we demonstrate molecular alignment under such conditions using a strong THz transient.

Figure 4 shows the retarded THz-induced adiabatic changes of the birefringence, observed on the time scale of tens of picoseconds, i.e. much longer than the stimulus pulse duration. The periodic change of birefringence originates from the off-resonant, THz-induced molecular alignment of the two principal air molecules $N_2$ and $O_2$ with quarter revival periods of $T_{N_2}^{1/4} \approx 2.1$ ps and $T_{O_2}^{1/4} \approx 2.9$ ps, respectively. The short THz pulse induces a transient dipole moment in the molecules which couples to the THz electric field. In classical terms, the laser electric field exerts a torque on the molecules leading to alignment along the polarization axis of the laser. The coherent superposition of rotational levels dephases at a rate proportional to the square of the wavepacket width in J-space. Since coherence is maintained the alignment revives at pre-determined times $T_{rev}=1/2Bc$ with B the rotational constant and c the speed of light. At half of the period (4.15 ps and 5.8 ps, respectively) the molecules are aligned perpendicular to the field. A maximum alignment is measured for the two first half-revivals ($N_2$). We observed revivals up to a delay of 120 ps, corresponding to the 15[th] (10[th]) revival for $N_2$ and $O_2$. Moreover, around a delay ~ 21 ps, we trace 1/2 revival period of $CO_2$ [34] superimposed on 5/2 period of $N_2$. This is the first observation of such periodic revivals induced by a low-frequency THz pulse in non-polar molecules. The non-ionizing lambda-cubic THz alignment scheme presented here is capable of providing a substantial degree of alignment of non-polar molecules.

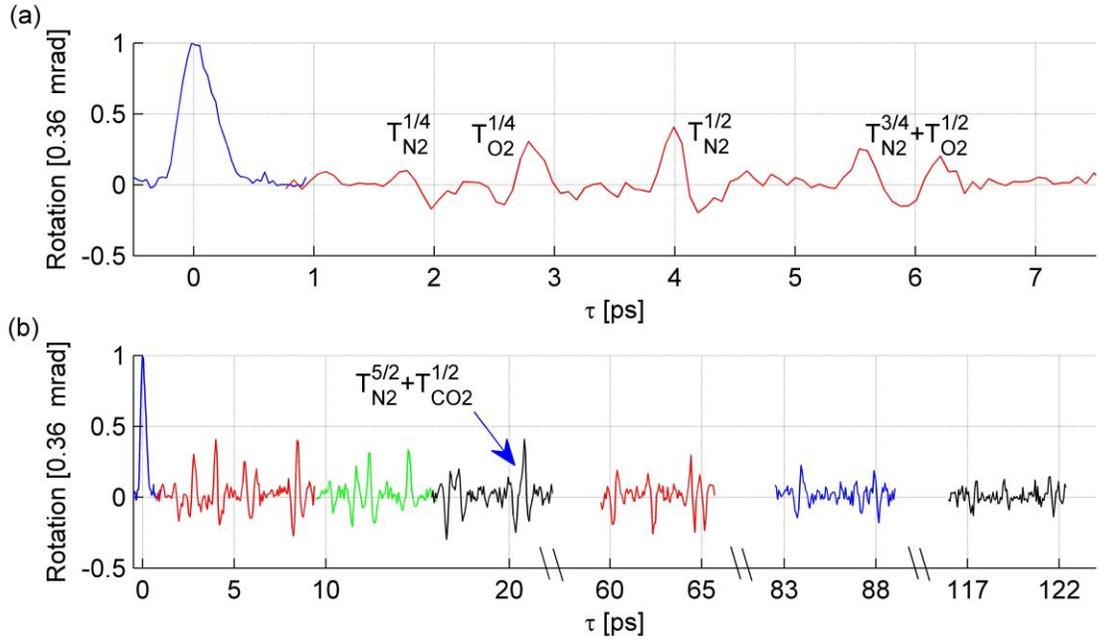

FIG. 4. (color online). (a) Induced birefringence initiated by the lambda-cubic THz pulse in air at ambient pressure and temperature. (b) Molecular alignment revivals for the non-polar $N_2$ and $O_2$ occur periodically at quarter-periods of 2.1 and 2.9 ps, respectively. At delay ~ 21 ps, a trace of 1/2 revival of $CO_2$ is shown superimposed on 5/4 revivals of the N2. At time zero the THz electric field transient induces a non-permanent dipole moment which couples to the THz single-cycle electric field component. As the system coherence is maintained the revivals were recorded up to 120 ps.

In conclusion, we presented the first demonstration of low-frequency nonlinearities in air, triggered by an ultra-intense lambda-cubic THz pulse. The experimental measurement of the instantaneous THz-induced Kerr nonlinearity coefficient $n_2$ will support future modeling and experimentation of nonlinear propagation of THz in air such as THz Kerr self-focusing, THz self-phase modulation and THz filamentation. On a delayed time scale, we observed THz-induced field-free alignment dynamics of non-polar air molecules. The revivals were initiated by a transient dipole moment induced by the strong THz pulse. Our approach represents a novel method for field-free alignment of non-polar molecules at intensities well below the ionization threshold which hold promise for enhanced control on molecular alignment.

**Acknowledgments:**

We are grateful to Marta Divall, Alexandre Trisorio, and Andeas Dax for operating the Ti:sapphire laser system. The OPA operation was supported by Carlo Vicario and Marta Divall. The data acquisition software, motors control, and THz camera used in this work were previously installed by Carlo Vicario, Clemens Ruchert, Rasmus Ischebeck, Edwin Divall, and Balazs Monoszlai. We acknowledge financial support from the Swiss National Science Foundation (SNSF) (grant no 200021_146769). MS acknowledges the support from a PSI-COFUND postdoctoral grant. CPH acknowledges association to NCCR-MUST and support from SNSF (grant no. PP00P2_150732).

[1] C. Marceau, Y. Chen, F. Théberge, M. Châteauneuf, J. Dubois, and S.L. Chin, Opt. Lett. **34**, 1417 (2009).

[2] C.P. Hauri, W. Kornelis, F.W. Helbing, A. Heinrich, A. Couairon, A. Mysyrowicz, J. Biegert, and U. Keller, Appl. Phys. B **79**, 673 (2004).

[3] T. Kampfrath, K. Tanaka, and K. A. Nelson, Nat. Photonics **7**, 680 (2013).

[4] C. Vicario, C. Ruchert, F. Ardana-Lamas, P. M. Derlet, B. Tudu, J. Luning, and C. P. Hauri, Nat. Photonics **7**, 720–723 (2013).

[5] M. Shalaby, J. Fabianska, M. Peccianti, Y. Ozturk, F. Vidal, H. Sigg, R. Morandotti, and T. Feurer, Appl. Phys. Lett. **104**, 171115 (2014).

[6] M. Liu, H. Y. Hwang, H. Tao, A. C. Strikwerda, K. Fan, G. R. Keiser, A. J. Sternbach, K. G. West, S. Kittiwatanakul, J. Lu, et al., Nature **487**, 345 (2012).

[7] M. Mochizuki, N. Nagaosa, Phys Rev. Lett. **105**, 147202 (2010).

[8] M. Shalaby, F. Vidal, M. Peccianti, R. Morandotti, F. Enderli, T. Feurer, and B. Patterson, Phys. Rev. B **88**, 140301(R) (2013).

[9] O. Schubert, M. Hohenleutner, F. Langer, B. Urbanek, C. Lange, U. Huttner, D. Golde, T. Meier, M. Kira, S. W. Koch, and R. Huber, Nat. Photonics **8**, 119 (2014).

[10] M. Machholm and N.E. Henriksen, Phys. Rev. Lett **78**, 193001 (2001).

[11] M. Lemeshko, R. V. Krems, J. M. Doyle, and S. Kais, Mol. Phys. **111**, 1648 (2013).

[12] R. Velotta, N. Hay, M. B. Mason, M. Castillejo, and J. P. Marangos, Phys. Rev. Lett. **87**, 183901 (2001).

[13] J. Itatani, J. Levesque, D. Zeidler, H. Niikura, H. Pepin, J. C. Kieffer, P. B. Corkum, and D. M. Villeneuve, Nature (London) **432**, 867 (2004).

[14] H. Soifer, P. Botheron, D. Shafir, A. Diner, O. Raz, B. D. Bruner, Y. Mairesse, B. Pons, and N. Dudovich, Phys. Rev. Lett. **105**, 143904 (2010).

[15] S. L.Chin, T.–J. Wang, C. Marceau, J. Wu, J. S. Liu, O. Kosareva, N. Panov, Y. P. Chen, J.–F. Daigle, S. Yuan, *et al.*, Laser Phys. **22**, 1 (2012).

[16] J. Wu, Y. Tong, M. Li, H. Pan, and H. Zeng, Phys. Rev. A **82**, 053416 (2010).

[17] J. Chem. Phys. **93**, 4779 (1990)

[18] J. Hajdu, Curr. Opin. Struct. Biol. **10**, 569 (2000).

[19] R. Neutze, R. Wouts, D. van der Spoel, E. Weckert, and J. Hajdu, Nature (London) **406**, 752 (2000).

[20] A. G. York, Opt. Express **17**, 13671 (2009).

[21] S. De, I. Znakovskaya, D. Ray, F. Anis, N. G. Johnson, I. A. Bocharova, M. Magrakvelidze, B. D. Esry, C. L. Cocke, I. V. Litvinyuk, and M. F. Kling, Phys. Rev. Lett. **103**, 153002 (2009); R. Tehini and D. Sugny, Phys. Rev. A **77**, 023407 (2008).

[22] H. Cai, J. Wu, Y. Peng, and H. Zeng, Opt. Express **17**, 5822 (2009); S. Varma, Y. H. Chen, and H. M. Milchberg, Phys. Rev. Lett. **101**, 205001 (2008); J. Wu, H. Cai, H. Zeng, and A. Couairon, Opt. Lett. **33**, 2593 (2008).

[23] M. Shalaby and C. P. Hauri, Nat. Commun. **6**, 5976 (2015).

[24] S. Fleischer, Y. Zhou, R. W. Field, and K. A. Nelson, Phys. Rev. Lett. **107**, 163603 (2011).

[25] K. N. Egodapitiya, S. Li, and R. R. Jones, Phys. Rev. Lett. **112**, 103002 (2014).

[26] H. Harde, S. Keiding, and D. Grischkowsky, Phys. Rev. Lett. **66**, 1834 (1991).

[27] V. Loriot, E. Hertz, O. Faucher, B. Lavorel, Opt. Express **17**, 13429 (2009).

[28] A. Yariv, *Optical Electronics in Modern Communications* (Oxford University Press, Oxford, 1997).

[29] R. W. Boyd, *Nonlinear Optics* (Academic, San Diego, Calif., 1992).

[30] M. Zalkovskij, A. C. Strikwerda, K. Iwaszczuk, A. Popescu, D. Savastru, R. Malureanu, A. V. Lavrinenko, and P. U. Jepsen, Appl. Phys. Lett. **103**, 221102 (2013).

[31] V. Loriot, E. Hertz, O. Faucher, B. Lavorel, Opt. Express **17**, 13429 (2009).

[32] M. Bache, F. Eilenberger, S. Minardi, Opt. Letters **37**, 4612 (2012).

[33] S. De, I. Znakovskaya, D. Ray, F. Anis, Nora G. Johnson, I. A. Bocharova, M. Magrakvelidze, B. D. Esry, C. L. Cocke, I. V. Litvinyuk, and M. F. Kling, Phys. Rev. Lett. **103**, 153002 (2009).

[34] G. H. Lee, H. T. Kim, J. Y. Park, C. H. Nam, T. K. Kim, J. H. Lee, and H. Ihee, J. Korean Phys. Soc. **49**, 337 (2006).